\documentclass[preprint,aps,a4paper,showpacs,amsfonts,amsmath,amssymb]{revtex4}

\usepackage{graphicx}

\begin{document}

\title{Extreme times for volatility processes}
\author{Jaume Masoliver and Josep Perell\'o}
\affiliation{Departament de F\'{\i}sica Fonamental, Universitat de
Barcelona.\\ Diagonal 647, E-08028 Barcelona, Spain}
\begin{abstract}
Extreme times techniques, generally applied to non-equilibrium statistical mechanical processes, are also useful for a better understanding of financial markets. We present a detailed study on the mean first-passage time for the volatility of return time series. The empirical results extracted from daily data of major indices seem to follow the same law regardless the kind of index thus suggesting an universal pattern. The empirical mean first-passage time to a certain level $L$ is fairly different from that of the Wiener process showing a dissimilar behavior depending on whether $L$ is higher or lower than the average volatility. All of this indicates a more complex dynamics in which a reverting force drives volatility toward its mean value. We thus present the mean first-passage time expressions of the most common stochastic volatility models whose approach is comparable to the random diffusion description. We discuss asymptotic approximations and confront them to empirical results with a good agreement with the ExpOU model.
\end{abstract}
\date{\today}
\maketitle

\section{Introduction}
\label{sec1}

First-passage and extreme value problems are a crucial aspect of stochastic methods with a long tradition of applications to physics, biology, chemistry and engineering, all of them related to non-equilibrium processes. We thus can mention driven granular matter or polymers passing through pores~\cite{koplik,redner,lamura}, chemical reactions dynamics~\cite{szabo}, reaction-diffusion systems~\cite{bray}, coarsening systems~\cite{tam,rapapa}, fluctuating interfaces~\cite{krug,majumdar} or even nuclear fussion and light emission~\cite{hofmann} among many others~\cite{refs,refs1,burkhardt,redner1}. In addition, many natural records need also a similar description as are, for instance, floods, very high temperatures and solar flares or earthquakes~\cite{eichner}. 

In studying extreme value statistics of a given time series one wants to learn about the distribution of the extreme events, that is, the maximum values of the signal within time intervals of fixed duration, and the statistical properties of their sequences. In hydrological engineering, for example, extreme value statistics are commonly applied to decide what building projects are required to protect riverside areas against typical floods that occur once in 100 years~\cite{eichner}. 

All this effort and knowledge have not substantially been introduced in the exploration of extreme events in financial markets and it has mostly remained inside physical sciences and engineering without any great spread outside them. Nevertheless, we believe that this perspective can be helpful by providing an alternative approach to extreme statistics that is different from that of the current mathematical finance~\cite{embrechts} which can also result in a better control of the risk in financial markets.

In the quantitative study of financial markets the volatility, originally defined as the standard deviation of returns, plays an increasingly important role as a measure of risk. There are nowadays many financial products, such as options and other financial derivatives, which are specifically designed to cover investors from the risk associated with any market activity. These products are fundamentally based on the volatility, therefore, its knowledge turns out to be essential in any modern financial setting and, hence, in any financial modeling. 

One of the earliest financial models, the model of Einstein-Bachelier~\cite{cootner}, assumes that the volatility is constant being itself the diffusion coefficient of a Wiener process. However, this assumption is questioned by many empirical observations which are gathered together in the so-called ``stylized facts''~\cite{cont}. The overall conclusion is that the volatility is not constant, it is not even a function of time but a random variable. Consequently the measure of volatility has become more difficult and questions like at what time the volatility reaches, for the first time, a determined value -- which may or may not be extreme-- are quite relevant. 

The main objective of this paper is therefore to study the mean first-passage time (MFPT) of the volatility process. We approach the problem both from analytical and empirical viewpoints. For one hand, we analyze the MFPT for daily data of major financial indices and observe that the MFPT curves of all indices follow an universal pattern when the volatility is scaled in a proper way. On the other hand, we obtain analytical expressions of the MFPT for a special class of two-dimensional diffusion models commonly known as stochastic volatility models in the quantitative finance literature~\cite{ronnie_book}. We next compare the analytical results with the empirical predictions which provides a test about the suitability of these analytical models. We incidentally note that these stochastic volatility models are analogous to the ones arising in the random diffusion framework~\cite{ben} and even to some multifractal models~\cite{saichev}.

As mentioned, previous works on extreme times are, to our knowledge, scarce and mostly dealing with the return process but not with volatility. In our early works~\cite{mmp,montero_lillo} we have analyzed the mean exit time of the return based on the continuous random walk technique and addressed exclusively to tick-by-tick data. Other examples studying the extreme time return statistics are given in Ref.~\cite{simonsen,jafari} and specially in Refs.~\cite{spagnolo,bonanno} where the MET for the stock price is simulated using an stochastic volatility model as underlying process. And finally, there are also recent studies focused on the volatility data analyzing the interevent time statistics between spikes~\cite{yamasaki,wang} or the survival probability comparing the high frequency empirics with results from multifractal modeling~\cite{constantin,saichev}.

We end this introductory section by pointing out that the analysis of extreme times is closely related to at least two challenging problems in mathematical finance which are of great practical interest: the American option pricing~\cite{kim,perello} and the issue of default times and credit risk~\cite{rutkowski,kijima}. Both problems require the knowledge of hitting times, that is, first-passage times to certain thresholds. However, the typical mathematical approach there is quite different from the one we study here.

The paper is organized as follows. In Sect.~\ref{sec2} we outline the most usual stochastic volatility models. In Sect.~\ref{sec3} we obtain the general expressions for the MFPT based on these models. In Sect.~\ref{sec4} we analyze the averaged extreme time and examine its asymptotic behavior. In Sect.~\ref{sec5} we estimate the empirical MFPT of several financial indices and compare it with the analytic expressions obtained in previous sections. Conclusions are drawn in Sect.~\ref{sec6} and some more technical details are left to Appendices.

\section{Stochastic volatility models}
\label{sec2}

The geometric Brownian motion (GBM) proposed by physicist Osborne in 1959~\cite{osborne} is, without any doubt, the most widely used model in finance. In this setting any speculative price $S(t)$ is described through the following Langevin equation (in It\^o sense)
\begin{equation}
\frac{dS(t)}{S(t)}=\mu dt+\sigma\ dW_1(t),
\label{1}
\end{equation}
where $\sigma$ is the volatility, assumed to be constant, $\mu$ is some deterministic drift indicating an eventual trend in the market, and $W_1(t)$ is the Wiener process. However, and particularly after the 1987 crash, there is a compelling empirical evidence that the assumption of constant volatility is doubtful~\cite{cont,grpi}, neither it is a deterministic function of time --as one might wonder on account of the non stationarity of financial data-- but a random variable. The volatility is often related to the market activity~\cite{luiggi}. In this way we are assuming that market activity is stochastic and governed by the random arrival of information to the markets. 

The hypothesis of a random volatility was originally suggested to explain the so-called ``smile effect'' appearing in the implied volatility of option prices~\cite{ronnie_book}. 
In the most general frame one therefore assumes that the volatility $\sigma$ is a given function of a random process $Y(t)$: 
\begin{equation}
\sigma(t)=\sigma(Y(t)).
\label{3}
\end{equation}

Most stochastic volatility (SV) models that have been proposed up till now suppose that $Y(t)$ is also a diffusion process that may or may not be correlated with price, and different models mainly differ from each other in the way that $\sigma$ depends on $Y(t)$. 

The usual starting point of the SV models is the GBM given by Eq.~(\ref{1}) with $\sigma$ given by Eq.~(\ref{3}) and $Y(t)$ being a diffusion process:
\begin{equation}
dY(t)=F(Y)dt+G(Y)dW_2(t).
\label{4}
\end{equation}
In Eqs.~(\ref{1}) and~({\ref{4}) $W_i(t)$ $(i=1,2)$ are Wiener processes, that is, $dW_i(t)=\xi_i(t)dt$, where $\xi_i(t)$ are zero-mean Gaussian white noises with $\langle\xi_i(t)\xi_i(t')\rangle=\delta(t-t')$ and cross correlation given by
\begin{equation}
\langle\xi_1(t)\xi_2(t')\rangle=r\delta(t-t')
\label{5a}
\end{equation}
$(-1\leq r \leq 1)$. Incidentally we note that any SV model defined through Eqs. (\ref{1}-\ref{5a}) is, in fact, a two-dimensional diffusion process. 

The most common SV models in the literature are the following:

(a) The {\it Ornstein-Uhlenbeck} (OU) model in which 
$$
\sigma=Y,\quad F(Y)=-\alpha(Y-m),\quad G(Y)=k;
$$
that is:
\begin{equation}
dY(t)=-\alpha(Y-m)dt+kdW_2(t). 
\label{5}
\end{equation}
See Refs.~\cite{Stein,perello2} for further details.

(b) The {\it Cox-Ingersoll-Ross-Heston} (CIR-Heston) model:
$$
\sigma=\sqrt{Y},\quad F(Y)=-\alpha(Y-m^2),\quad G(Y)=k\sqrt{Y};
$$
then
\begin{equation}
dY(t)=-\alpha (Y-m^2)dt+k\sqrt{Y}dW_2(t).
\label{7}
\end{equation}
See Refs.~\cite{cox,heston,yakov} for further details.

(c) The {\it exponential Ornstein-Uhlenbeck} (ExpOU) model:
$$
\sigma=me^{Y},\quad F(Y)=-\alpha Y, \quad G(Y)=k;
$$
hence
\begin{equation}
dY(t)=-\alpha Y dt+k dW_2(t). 
\label{9}
\end{equation}
See Refs.~\cite{fouque,perello3} for further details.

From the above equations we see that the volatility is also described by a one-dimensional diffusion process:
\begin{equation}
d\sigma(t)=f(\sigma)dt+g(\sigma)dW(t).
\label{3-1}
\end{equation}
Thus, for the OU model $\sigma=Y$ and (see Eq.~(\ref{5})):
\begin{equation}
d\sigma(t)=-\alpha(\sigma-m)dt+kdW(t).
\label{3-2}
\end{equation}
However, obtaining a differential equation for $\sigma(t)$ for CIR-Heston and ExpOU models is not direct, since in these cases the volatility, $\sigma=\sigma(Y)$, is a nonlinear function of processes $Y$ and the differentials of $\sigma$ and $Y$ are connected through the It\^o lemma~\cite{perello-masoliver,gardiner}:
\begin{equation}
d\sigma(Y)=\frac{\partial\sigma}{\partial Y}dY+\frac{1}{2}\frac{\partial^2\sigma}{\partial Y^2}dt.
\label{ito}
\end{equation}
For the CIR-Heston model $\sigma=\sqrt{Y}$ and from Eqs. ~(\ref{7}) and ~(\ref{ito}) we get
\begin{equation}
d\sigma(t)=-\frac{1}{2}\alpha\left(\sigma-\frac{\rho}{\sigma}\right)+kdW(t),
\label{3-3}
\end{equation}
where
\begin{equation}
\rho\equiv m^2-\frac{k^2}{4\alpha}.
\label{rho}
\end{equation}
In the case of the ExpOU model $\sigma=me^{Y}$, and
\begin{equation}
d\sigma(t)=-\alpha\ln(\sigma/M)dt+k\sigma dW(t),
\label{3-4}
\end{equation}
where
\begin{equation}
M\equiv me^{k^2/2\alpha}.
\label{M}
\end{equation}

\section{The mean first-passage time}
\label{sec3}

In this section and the next, we study the MFPT of the volatility process from an analytical point of view. We postpone for a later section, Sect.~\ref{sec5}, the analysis of empirical mean first-passage times for several markets and their comparison with the analytical expressions obtained in Sects.~\ref{sec3}-\ref{sec4}. 

Let us denote by $T_\lambda(\sigma)$ the MFPT of the volatility process. That is, $T_\lambda(\sigma)$ represents the mean time one has to wait in order to observe the volatility, starting from a known value $\sigma$, to reach for the first time a prescribed value $\lambda$, which we often refer to as the ``critical level''. 

If we assume that the volatility is given by the diffusion process described in Eq.~(\ref{3-1}), then $T_\lambda(\sigma)$ obeys the following differential equation~\cite{gardiner} 
\begin{equation}
\frac{1}{2}g^2(\sigma)\frac{d^2T_\lambda}{d\sigma^2}+f(\sigma)\frac{dT_\lambda}{d\sigma}=-1
\label{3-5}
\end{equation}
with an absorbing boundary condition at the critical level:
\begin{equation}
T_\lambda(\lambda)=0.
\label{3-6}
\end{equation}
Let us recall that the volatility should be a positive defined quantity. Hence, we have also to impose a reflection when it reaches the value $\sigma=0$. This is achieved by adding the reflecting boundary condition:
\begin{equation}
\left.\frac{dT_\lambda}{d\sigma}\right|_{\sigma=0}=0.
\label{3-7}
\end{equation}

Before proceeding ahead we note that the most general approach to the problem at hand would be obtaining the MFPT, $T(S,\sigma)$, of the two-dimensional process $(S(t),\sigma(t))$ defined in Eqs. (\ref{1})-(\ref{4}). After knowing $T(S,\sigma)$ --certainly a difficult mathematical task-- we can get two different extreme times. Thus by averaging the volatility out of $T(S,\sigma)$ we have $T(S)$, {\it i.e.}, the MFPT for the price. On the other hand, averaging the price out of $T(S,\sigma)$ one gets the MFPT for the volatility $T(\sigma)$, the latter being the main objective of the present work. Obviously $T(\sigma)$ is much easier to obtain from Eqs. (\ref{3-5})-(\ref{3-7}) than from this general and somewhat tortuous proceeding based on $T(S,\sigma)$. 

Let us return to the solution of the problem posed by Eqs.~(\ref{3-5})--(\ref{3-7}). This can be attained by elementary means with the result
\begin{equation}
T_\lambda(\sigma)=2\int_\sigma^\lambda e^{-\psi(x)}dx\int_0^x\frac{e^{\psi(y)}}{g^2(y)}dy,
\label{3-8}
\end{equation}
where
\begin{equation}
\psi(x)=2\int\frac{f(x)}{g^2(x)}dx.
\label{3-9}
\end{equation}
We shall now evaluate the expressions taken by the MFPT, $T_\lambda(\sigma)$, according to the SV model chosen.

(a) The {\it OU model}. In this case (see Eq.~(\ref{3-2})) 
$$
f(x)=-\alpha(x-m), \qquad g(x)=k.
$$
Hence
\begin{equation}
\psi(x)=-\nu^2(x^2-2mx),
\label{3-10}
\end{equation}
where
\begin{equation}
\nu\equiv\frac{\alpha^{1/2}}{k}.
\label{3-11}
\end{equation}
Using Eq.~(\ref{3-8}) with Eq.~(\ref{3-10}), we get
\begin{equation}
T_\lambda(\sigma)=\frac{2}{\alpha}\int_{\nu(\sigma-m)}^{\nu(\lambda-m)}e^{x^2}dx\int_{-\nu m}^xe^{-y^2}dy
\label{3-12}
\end{equation}
which in turn can be written as
\begin{equation}
T_\lambda(\sigma)=\frac{\sqrt{\pi}}{\alpha}\int_{\nu(\sigma-m)}^{\nu(\lambda-m)}e^{x^2}\left[{\rm erf}(\nu m)+{\rm erf}(x)\right]dx,
\label{3-13}
\end{equation}
where
$$
{\rm erf}(x)=\frac{2}{\sqrt{\pi}}\int_0^xe^{-y^2}dy
$$
is the error function. 

(b) The {\it CIR-Heston model}. From Eq.~(\ref{3-3}) we have 
$$
f(x)=\frac{1}{2}\alpha\left(x-\frac{\rho}{x}\right), \qquad g(x)=k,
$$
and
\begin{equation}
\psi(x)=-2\nu^2 x^2+4m\nu^2\ln x.
\label{3-14}
\end{equation}
After substituting Eq.~(\ref{3-14}) into Eq.~(\ref{3-8}), some elementary manipulations yield
\begin{equation}
T_\lambda(\sigma)=\frac{1}{\alpha}\int_{2\nu^2\sigma^2}^{2\nu^2\lambda^2}x^{-\beta}e^{x}dx\int_{0}^xy^{\beta-1}e^{-y}dy,
\label{3-15}
\end{equation}
where
\begin{equation}
\beta\equiv\frac{1}{2}(1+4\rho\nu^2).
\label{3-16}
\end{equation}
We use the following integral representation of the confluent hypergeometric function~\cite{MOS}
\begin{widetext}
$$
F(a,c,x)=\frac{\Gamma(c)}{\Gamma(a)\Gamma(c-a)}e^{-x}x^{a-c}\int_0^xe^{-y}y^{c-a-1}(1-xy)^{a-1}dy,
$$
\end{widetext}
and write Eq.~(\ref{3-15}) in its final form
\begin{equation}
T_\lambda(\sigma)=\frac{1}{\alpha\beta}\int_{2\nu^2\sigma^2}^{2\nu^2\lambda^2}F(1,1+\beta,x)dx.
\label{3-17}
\end{equation}

(c) The {\it ExpOU model}. In this case ({\it cf.} Eq.~(\ref{3-4})) 
$$
f(x)=-\alpha x\ln(x/M), \qquad g(x)=kx.
$$
Consequently
\begin{equation}
\psi(x)=-\nu^2\ln^2(x/M),
\label{3-18}
\end{equation}
where $\nu$ and $M$ are given by Eqs.~(\ref{3-11}) and~(\ref{M}) respectively. As before, substituting Eq.~(\ref{3-18}) into Eq.~(\ref{3-8}) and some elementary manipulations involving simple change of variables inside the integrals, result into
\begin{equation}
T_\lambda(\sigma)=\frac{2}{\alpha}\int_{\eta(\sigma)}^{\eta(\lambda)}e^{x^2}dx\int_{-\infty}^xe^{-y^2}dy,
\label{3-19}
\end{equation}
where
\begin{equation}
\eta(\sigma)\equiv\frac{1}{2\nu}+\nu\ln(\sigma/m).
\label{3-20}
\end{equation}
Note that using the error function defined above we can write $T_\lambda(\sigma)$ up to a quadrature by
\begin{equation}
T_\lambda(\sigma)=\frac{\sqrt{\pi}}{\alpha}\int_{\eta(\sigma)}^{\eta(\lambda)}e^{x^2}\left[1+{\rm erf}(x)\right]dx.
\label{3-21}
\end{equation}

We finish this section reminding what is the MFPT for the simplest SV model. This is the case when the volatility is totally random without any reverting force driving the volatility to its normal level, that is
\begin{equation}
d\sigma(t)=kdW(t),
\label{dx}
\end{equation}
where $k$ is a constant. We remark that, in contrast with the previous SV models, this dynamics has no temporal correlations at all, {\it i.e.,} it has no memory. To compute its MFPT, we take again Eqs.~(\ref{3-8})-(\ref{3-9}). We insist in the fact the we are assuming a reflecting barrier for $\sigma=0$ in order to get a dynamics restricted between 0 and $\lambda$. This is in fact the reason why we obtain a finite MFPT since without a reflecting barrier which prevents $\sigma$ to be negative the MFPT would not exist~\cite{gardiner}. In such a case, after including the reflecting barrier at $\sigma=0$, it is straightforward to get
\begin{equation}
T_\lambda(\sigma)=\frac{1}{k^2}(\lambda^2-\sigma^2).
\label{random}
\end{equation}
This will be our benchmark solution in future sections. 

\section{The averaged MFPT}
\label{sec4}

The expressions for $T_\lambda(\sigma)$ developed in the previous section give us the mean time one has to wait until the volatility reaches a given level $\lambda$ starting from its present value $\sigma$. However, it is also of theoretical and practical interest~\cite{mmp,montero_lillo,refs1} the knowledge of the {\it averaged MFPT} in which the dependence on $\sigma$ has been averaged out. One might argue that this quantity has fewer applications to trading and investment but, as we will see later, for the current purposes of this paper this simplification really helps to reach relevant conclusions based on real data. 

To obtain this average we have to choose a probability distribution for $\sigma$. The simplest and most common assumption takes the present value of the volatility as uniformly distributed over the interval $(0,\lambda)$. For our purposes this choice is also very convenient because the average performed is independent of the SV model chosen, {\it i.e.}, it is the same average for all models. We will thus test their abilities to reproduce real data under identical conditions. 

We thus define
\begin{equation}
\overline{T}(\lambda)\equiv\frac{1}{\lambda}\int_0^\lambda T_\lambda(\sigma)d\sigma.
\label{4-1}
\end{equation}
We can easily obtain $\overline{T}(\lambda)$ for the SV models discussed above. This is done at once by substituting into Eq.~(\ref{4-1}) the expressions of $T_\lambda(\sigma)$ given by those models. For the OU, CIR-Heston and ExpOU models ({\it cf.} Eqs.~(\ref{3-13}),~(\ref{3-17}) and~(\ref{3-21})) this replacement, followed by an integration by parts, yields respectively
\begin{widetext}
\begin{equation}
\overline{T}(\lambda)=\frac{\nu\sqrt{\pi}}{\alpha\lambda}\int_{0}^{\lambda}xe^{\nu^2(x-m)^2}
\bigl[{\rm erf}(\nu m)+{\rm erf}(\nu(x-m))\bigr]dx,
\label{4-2}
\end{equation}
\end{widetext}
\begin{equation}
\overline{T}(\lambda)=\frac{2^{-1/2}}{\alpha\beta\nu\lambda}\int_{0}^{2\nu^2\lambda^2}x^{1/2}F(1,1+\beta,x)dx,
\label{4-3}
\end{equation}
and
\begin{equation}
\overline{T}(\lambda)=\frac{m\sqrt{\pi}}{\alpha\lambda}\int_{-\infty}^{\eta(\lambda)}e^{x^2+x/\nu}\left[1+{\rm erf}(x)\right]dx,
\label{4-4}
\end{equation}
where, in writing the last equation we have used the definition of the parameter $M$ given in Eq.~(\ref{M}). For the memoryless random volatility given by Eqs.~(\ref{dx})-(\ref{random}) we have
\begin{equation}
\overline{T}(\lambda)=\frac{2\lambda^2}{3k^2}.
\label{4-5b}
\end{equation}

Equations~(\ref{4-2}) and~(\ref{4-3}) are the final expressions of the averaged MFPT for the OU and CIR-Heston models. The expression given by Eq.~(\ref{4-4}), corresponding to the ExpOU model can be written in an alternative form which will show their usefulness both in the asymptotic and empirical analyses to be undertaken below. Thus, in Eq.~(\ref{4-4}) we change the variable $x \rightarrow -x$ and use the identity 
${\rm erfc} (x)=1-{\rm erf} (x)$ (${\rm erfc}(x)$ is the complementary error function). We have
$$
\overline{T}(\lambda)=\frac{m\sqrt{\pi}}{\alpha\lambda}\int_{-\eta(\lambda)}^{\infty}e^{x^2-x/\nu}{\rm erfc}(x)dx.
$$
The complementary error function can be written in terms of the Kummer function $U(a,c,x)$ as~\cite{MOS}
$$
{\rm erfc}(x)=\frac{1}{\sqrt{\pi}}e^{-x^2}U\left(\frac{1}{2},\frac{1}{2},x^2\right).
$$
Therefore,
\begin{equation}
\overline{T}(\lambda)=\frac{m}{\alpha\lambda}\int_{-\eta(\lambda)}^{\infty}e^{-x/\nu}U\left(\frac{1}{2},\frac{1}{2},x^2\right)dx,
\label{4-5}
\end{equation}
which is our final expression of the averaged MFPT corresponding to the ExpOU model.

\subsection{Scaling the critical level}

Before proceeding with the asymptotic analysis of the above expressions of $\overline{T}(\lambda)$, it is convenient to scale $\lambda$ so as to render it dimensionless. This will also be useful for the empirical analysis of the next section. 

In finance, it is rather relevant the knowledge of the so-called ``volatility's normal level'', $\sigma_{\rm s}$, which is defined as the mean stationary value of the volatility process:
\begin{equation}
\sigma_{\rm s}=\int_{-\infty}^{\infty}\sigma p_{\rm st}(\sigma)d\sigma,
\label{4-6}
\end{equation}
where $p_{\rm st}(\sigma)$ is the stationary probability density function (pdf) of the volatility random process. The significance of $\sigma_{\rm s}$ lies in the empirical fact that the volatility is, as time increases, {\it mean reverting} to its normal level~\cite{cont}. 

We therefore scale the critical level $\lambda$ with the normal level, and define
\begin{equation}
L\equiv\frac{\lambda}{\sigma_{\rm s}}.
\label{4-7}
\end{equation}
From the analytical view, the dimensionless critical value $L$ depends on the SV model we choose. The calculation of $\sigma_{\rm s}$ for different SV models is given in Appendix~\ref{appendixA} with the result:
\begin{equation}
\sigma_{\rm s}=\begin{cases} m & \text{OU model,} \\
 \gamma/\nu\sqrt{2} & \text{CIR-Heston model,} \\
 me^{1/4\nu^2} & \text{ExpOU model,}
 \end{cases}
\label{normal-level} 
\end{equation} 
where
\begin{equation}
\gamma=\frac{\Gamma(\beta+1/2)}{\Gamma(\beta)},
\label{4-11}
\end{equation}
and $\beta$ is defined in Eq. (\ref{3-16}). 

Once we know the normal level, substituting $\lambda$ in terms of $L$ into Eqs.~(\ref{4-2}),~(\ref{4-3}) and~(\ref{4-5}) result in the following expressions for the averaged MFPT:

(a) The {\it OU model}
\begin{widetext}
\begin{equation}
\overline{T}(L)=\frac{m\nu \sqrt{\pi}}{\alpha L}\int_{0}^{L}xe^{m^2\nu^2(x-1)^2}\bigl[{\rm erf}(m\nu)
+{\rm erf}(m\nu(x-1))\bigr]dx.
\label{4-8}
\end{equation}
\end{widetext}

(b) The {\it CIR-Heston model}
\begin{equation}
\overline{T}(L)=\frac{1}{\alpha\beta\gamma L}\int_{0}^{\gamma^2 L^2}x^{1/2}F(1,1+\beta,x)dx,
\label{4-10}
\end{equation}
where $\nu$ and $\gamma$ are respectively defined in Eqs.~(\ref{3-11}) and~(\ref{4-11}).

(c) The {\it ExpOU model}
\begin{equation}
\overline{T}(L)=\frac{e^{-1/4\nu^2}}{\alpha L}\int_{-\zeta(L)}^{\infty}e^{-x/\nu}U\left(\frac{1}{2},\frac{1}{2},x^2\right)dx,
\label{4-12}
\end{equation}
where
\begin{equation}
\zeta(L)\equiv \frac{1}{4\nu}+\nu\ln L.
\label{4-13}
\end{equation}

Finally, we observe that the scaling of the critical level through the normal level $\sigma_{\rm s}$ given in Eq. (\ref{4-7}) cannot be performed on the memoryless model (\ref{dx}). In this situation the volatility never reaches a stationary state and, hence, $\sigma_{\rm s}$ is meaningless. If however we define a new critical level as $L=\lambda/k$ (now $L$ has dimension of time) we can write 
\begin{equation}
\overline{T}(L)=\frac{2}{3}L^2.
\label{4-14}
\end{equation}

\subsection{Asymptotic analysis of the MFPT for small values of the critical level}

We will now obtain the asymptotic behavior of the averaged MFPT for small values of $L$. Note that the case 
$L\ll 1$ is equivalent to $\lambda\ll\sigma_{\rm s}$, that is, the critical level is a small fraction of the normal level.

For the OU model $\overline{T}(L)$ is given by Eq.~(\ref{4-8}) and one can easily see by means of a direct expansion that
\begin{widetext}
\begin{equation}
\int_{0}^{L}xe^{m^2\nu^2(x-1)^2}[{\rm erf}(m\nu)
+{\rm erf}(m\nu(x-1))]dx = \frac{2m\nu}{3\sqrt{\pi}}L^3+O\left(L^4\right).
\end{equation}
\end{widetext}
Hence 
\begin{equation}
\overline{T}(L)=\frac{2m^2}{3k^2}L^2+O\left(L^3\right).
\label{4-14a}
\end{equation}

For the CIR-Heston Model, Eq.~(\ref{4-10}), we use the expansion~\cite{MOS}
$$
F(1,1+\beta,x)=\sum_{n=0}^\infty\frac{x^n}{(1+\beta)_n},
$$
and get
\begin{equation}
\overline{T}(L)=\frac{2\gamma^2}{3\alpha\beta}L^2+O\left(L^4\right).
\label{4-15}
\end{equation}

The case of the ExpOU model requires a different approach than that of direct expansions. We first note that when 
$L\rightarrow 0$ the function $\zeta(L)$ defined in Eq.~(\ref{4-13}) tends to $-\infty$. Hence, in this limit the argument of the Kummer function $U(1/2,1/2,x^2)$ appearing in the integrand of Eq.~(\ref{4-12}) is exceedingly large. We can thus use the approximation~\cite{MOS}
$$
U\left(\frac{1}{2},\frac{1}{2},x^2\right)\simeq \frac{1}{x}+O\left(\frac{1}{x^3}\right),
$$
with the result 
$$
\overline{T}(L)\simeq\frac{e^{-1/4\nu^2}}{\alpha L}{\rm E}_1\left[-\zeta(L)/\nu\right],
$$
where 
$$
{E}_1(x)=\int_x^\infty\frac{e^{-t}}{t}dt
$$
is the exponential integral. Using the asymptotic approximation~\cite{MOS}
$$
{\rm E}_1(x)\simeq\frac{e^{-x}}{x}\left[1+\frac{1}{x}\right],
$$
and taking into account ({\it cf.} Eq.~(\ref{4-13})) 
$$
\frac{\zeta(L)}{\nu}=\frac{1}{4\nu^2}+\ln L \simeq \ln L, \qquad(L\ll 1);
$$
we finally obtain
\begin{equation}
\overline{T}(L)\simeq-\frac{1}{\alpha \ln L}+O\left(\frac{1}{\ln^2L}\right).
\label{4-16}
\end{equation}

Note that $\overline{T}(L)$ behaves, as $L\rightarrow 0$, in different ways depending on the SV model chosen. While in OU and CIR-Heston models (and in the memoryless model as well) the averaged MFPT grows quadratically with $L$, in the ExpOU model it grows logarithmically --an analogous situation arises for large values of $L$, see below. We will return to this point in the following sections. 

\subsection{Asymptotic analysis of the MFPT for large values of the critical level}

Let us now address the case $L\gg 1$ when the critical level is much greater than the normal level. In the Appendix~\ref{appendixB} we show that an asymptotic representation of $\overline{T}(\lambda)$ for large values of $\lambda$ and regardless the SV model chosen it is given by
\begin{equation}
\overline{T}(\lambda)\sim\frac{\lambda\sqrt{\pi}}{g^2(x_m)\sqrt{-2\psi''(x_m)}}e^{-\psi(\lambda)}.
\label{4-22}
\end{equation}
where $\psi(x)$ is defined in Eq.~(\ref{3-9}), and $x_m$ is the location of the maximum of $\psi(x)$. The values of $x_m$ are 
(see Appendix~\ref{appendixB}) 
\begin{equation}
x_m=\begin{cases} m & \text{OU model,}\\
 \sqrt{\rho} & \text{CIR-Heston model,}\\
 M & \text{ ExpOU model}, 
 \end{cases}
\label{4-20}
\end{equation}
where $\rho$ and $M$ are given by Eqs.~(\ref{rho}) and~(\ref{M}) respectively. 

For the SV models we are dealing with, Eq.~(\ref{4-22}) up to the leading order of $L$ yields (we use dimensionless units, {\it cf.} Eqs.~(\ref{4-7})-~(\ref{4-11})):

(a) {\it OU model}
\begin{equation}
\overline{T}(L)\sim\frac{m\sqrt{\pi}}{2\nu k^2}Le^{m^2\nu^2 L^2},
\label{4-23}
\end{equation}
where where $\nu$ is defined in Eq.~(\ref{3-11}). 

(b) {\it CIR-Heston model}
\begin{equation}
\overline{T}(L)\sim\frac{\gamma\sqrt{\pi/2}}{4\alpha}Le^{\gamma^2 L^2},
\label{4-24}
\end{equation}
where $\gamma$ is defined in Eq.~(\ref{4-11}).

(c) {\it ExpOU model}
\begin{equation}
\overline{T}(L)\sim\frac{\sqrt{\pi}}{2\nu k^2}Le^{\nu^2\ln^2L},
\label{4-25}
\end{equation}
where we have assumed that $-1/(4\nu^2)+\ln L \simeq \ln L$ for $L\gg 1$.

\section{Empirical results}
\label{sec5}

\begin{table*}
\begin{tabular}{|c|c|c|c|} \hline
Financial Indices & Period & data points & normal level (days$^{-1/2}$) \\
\hline 
DJIA & 1900-2004 & 28,545 & $7.1\times 10^{-3}$ \\
S\&P-500 & 1943-2003 & 15,152 & $6.2\times 10^{-3}$ \\
DAX & 1959-2003 & 11,024 & $8.4\times 10^{-3}$ \\
NIKKEI & 1970-2003 & 8,359 & $9.6\times 10^{-3}$ \\
NASDAQ & 1971-2004 & 8,359 & $7.8\times 10^{-3}$ \\
FTSE-100 & 1984-2004 & 5,191 & $7.7\times 10^{-3}$ \\
IBEX-35 & 1987-2004 & 4,375 & $9.6\times 10^{-3}$ \\
CAC-40 & 1987-2003 & 4,100 & $10.2\times 10^{-3}$ \\
\hline
\end{tabular}
\caption{\label{Table 1} Empirical data used}
\label{table1}
\end{table*}

We now present an empirical study of the MFPT for the daily volatility of major financial indices: (1) Dow-Jones Industrial Average (DJIA). (2) Standard and Poor's-500 (S\&P-500). (3) German index DAX. (4) Japanese index NIKKEI. (5) American index NASDAQ. (6) British index FTSE-100. (7) Spanish index IBEX-35. (8) French index CAC-40 (see Table~\ref{table1} for more details).

The empirical MFPT that we will obtain is the averaged MFPT and we will see its behavior in terms of the critical value. Let us recall that the uniform average has the advantatge over other more elaborated averages --as, for instance, averaging over the stationary distribution of the volatility-- that the former is independent of the SV model chosen which allows us to look at data on an identical footing when we confront the empirical observations with the predictions of any theoretical model. 

We work with the dimensionless level $L$ defined in Eq.~(\ref{4-7}) so that we first need to know from data the volatility's normal level, $\sigma_{\rm s}$, of each market (see Table~\ref{table1} for the empirical values of $\sigma_{\rm s}$). In this way we deal with critical levels that are proportional to the specific normal level of every market, thus unifying the magnitudes involved in the MFPT computation.

We incidentally note that instead of the averaged MFPT $\overline{T}(\lambda)$ we could have dealt with $T_\lambda(\sigma)$, that is, the MFPT starting from an specific value $\sigma$ and whose general expression is given in Eq. (\ref{3-8}). However, as we have observed already in this case the statistics of data becomes less reliable. Moreover the casuistry in the data analysis become more complex and disentangling all their properties goes far away from our main objectives. 

\subsection{Measuring the volatility}

From the time series of all the indices shown in Table~\ref{table1} we have to extract first the daily volatility and then evaluate the averaged MFPT for many values of $L$. Before proceeding further we observe that, in fact, {\it the volatility is never directly observed}. In practice, one usually takes as an approximate measure of the instantaneous volatility (over the time step $\Delta t$ assumed to be one day in our case) the quantity 
\begin{equation}
\sigma(t)\sim\frac{|\Delta X(t)|}{\sqrt{\Delta t}},
\label{5-1}
\end{equation}
where $\Delta X(t)$ is the daily zero-mean return change defined as follows 
\begin{equation}
\Delta X(t)=\frac{\Delta S(t)}{S(t)}-E\left[\frac{\Delta S(t)}{S(t)}\Bigg|S(t)\right],
\label{5-1a}
\end{equation}
where $\Delta S(t)=S(t+\Delta t)-S(t)$ are the daily price changes. The expected value appearing in this expression represents the conditional average of the relative price change knowing the current price $S(t)$. 

We will now justify the measure of $\sigma(t)$ given by Eq.~(\ref{5-1}). Using Eq.~(\ref{1}) as the evolution equation of $S(t)$, one can easily see that Eq.~(\ref{5-1a}) yields~\cite{footnote1}
\begin{equation}
\Delta X(t)\simeq\sigma(t) \Delta W(t).
\label{5-1b}
\end{equation}
Consequently $|\Delta X(t)|\simeq|\sigma(t)||\Delta W(t)|$; but $|\sigma(t)|=\sigma(t)$ and 
$$
|\Delta W(t)|=\sqrt{\Delta W(t)^2}\sim \Delta t,
$$
where the last expression must be understood in mean-square sense since $\Delta W(t)^2\rightarrow\Delta t$ for sufficiently small $\Delta t$~\cite{gardiner}. Collecting results we obtain 
$$
|\Delta X(t)|\sim\sigma(t)\Delta t
$$
which is Eq.~(\ref{5-1})~\cite{footnote2}.

From the extreme-time problem point of view, we can check the soundness of this procedure ({\it i.e.}, identifying $\sigma(t)$ by $|\Delta X(t)|/\Delta t$) comparing the results given by the exact analytical expressions of the MFPT for $\sigma$ obtained in Sec. \ref{sec4} with computer simulations of $|\Delta X(t)|$. Unfortunately the evaluation of the MFPT ${\overline T}(L)$, for any critical lavel $L$, has to be performed by the numerical integration of Eqs. (\ref{4-8})-(\ref{4-12}). The parameters we use for the numerical calculations are those given in the literature to reproduce the DJIA~\cite{perello2,yakov,perello3} and they are summarized in Table \ref{table2}.
\begin{table*}
\begin{tabular}{|l|c|c|c|c|} \hline
Parameters & $k$ & $\alpha$ & $m$ & $r$ \\
\hline
Ornstein-Uhlenbeck (OU) & $1.4\times 10^{-3}$ & $5\times 10^{-2}$ & $1.2\times 10^{-2}$ & -0.4 \\
Heston & $2.45\times 10^{-3}$ & $4.5\times 10^{-2}$ & $9.28\times 10^{-3}$ & -0.4 \\
Exponential OU (ExpOU) & $4.7\times 10^{-2}$ & $1.82\times 10^{-3}$ & $8\times 10^{-3}$ & -0.4 \\
\hline
\end{tabular}
\caption{The parameters, measured in daily units, for the SV models defined in Eqs.~(\ref{1})--(\ref{9}).}
\label{table2}
\end{table*}
The methodology for evaluating the MFPT for $|\Delta X|$ from simulated time series is the same as that of the empirical data oulined few lines below. One should mention that the numerical computation of the integral is quite fast and straightforward except for the expOU model with a difficult numerical convergence from small values of $L$ until the level raises the normal level of volatility. 

In Figure~\ref{fig0} we compare the two different results of the MFPT depending on the way we measure the volatility. The numerical form for $\sigma(t)$ may differ significantly from the simulation of $|\Delta X|$. In all cases we observe that the simulated $|\Delta X|$ is noisier than $\sigma(t)$ as should have been expected because of the extra noise source $\Delta W(t)$ appearing in Eq. (\ref{5-1b}). Thus the noise $\Delta W$ brings $|\Delta X|\simeq
\sigma |\Delta W|$ to both larger and smaller values than those typically given by $\sigma$ alone. However, in the asymptotic limit the functional forms of $|\Delta X|$ and $\sigma$ are completely similar.

\begin{figure}
\begin{center}
\includegraphics[width=0.45\textwidth,keepaspectratio=true]{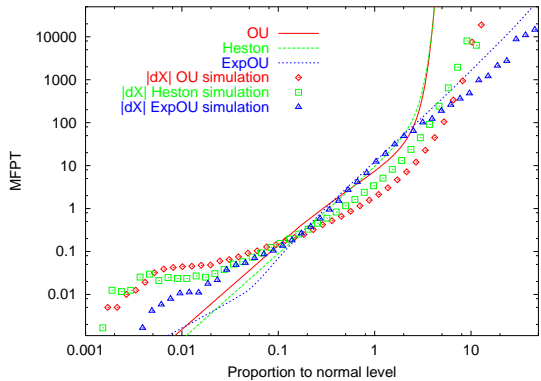} 
\caption{Log-log representation of the MFPT for the OU, CIR-Heston and expOU models. Points are the MFPT results for $|\Delta X(t)|$ sample paths of 100,000 steps. The curves correspond to the numerical computation of the MFPT for $\sigma$ using the three integrals given by Eqs.~(\ref{4-8})--(\ref{4-12}). Differences indeed appear but the qualitative behavior is preserved.}
\label{fig0}
\end{center}
\end{figure}

Before going further, one should mention that the procedure just outlined to catch the true volatility $\sigma$ is not unique. There is, at least, another relatively simple way of extracting it from the time series of prices~\cite{perello3}. This alternative method basically consists in dividing the empirical $\Delta X(t)$ --obtained through Eq.~(\ref{5-1a})-- by a simulated Gaussian process replicating the Wiener increments $\Delta W(t)$. This, after using Eq.~(\ref{5-1b}), yields an empirical value for $\sigma(t)$. We have shown in~\cite{perello3} that such a ``deconvolution procedure'' works relatively well and it reproduces reliable values of the empirical volatility as long as the effects of memory and cross-correlation are negligible (or, at least, that they do not affect the statistical analysis). However, in the analysis of extreme times the deconvolution method may destroy many subtleties of the MFPT curves. In this case, as we will see below, the market memory seems to really matter.

\subsection{The empirical MFPT}

Once we have a volatility time series constructed using Eq.~(\ref{5-1}), we can compute the mean first-passage time~\cite{footnote3}. In Fig.~\ref{fig1} we present the log-log representation of the empirical $\overline{T}(L)$ in terms of $L$ for the markets outlined in Table~\ref{table1}. We observe that the MFPT behaves in a very similar way for all markets and universality is well-sustained. In the same plot we have added several curves. The solid one gives the MFPT for the volatility of the (memoryless) Brownian motion process: $d\sigma=kdW(t)$. In this case, as we have shown in Eq. (\ref{4-14}), the averaged MFPT is a quadratic function of $L$, $\overline{T}(L)=(2/3)L^2$, for all $L>0$. Clearly, this behavior is not observed in real data. It seems to be therefore necessary to include a non uniform driving force in the drift that provides long correlations and clustering to the volatility dynamics. Moreover, we can fit two power laws of the form 
\begin{equation}
\overline{T}(L)\sim L^\alpha,
\label{power-law}
\end{equation}
with a different exponent depending on whether $L< 1$ or $L> 1$ ({\it cf} Fig.~\ref{fig1}). 

As we have just remarked the quadratic law $\overline{T}(L)\propto L^2$ corresponds to a purely random volatility without memory. Moreover, for the standard GBM given by $dX=\sigma dW$, where $X(t)$ is the zero-mean return and the volatility is constant, we see at once that the averaged MFPT for $|\Delta X|$ obeys the same law:
\begin{equation}
\overline{T}(L)\propto L^2.
\label{power-law(a)}
\end{equation}
Therefore, following the discussion of the previous section about using $|\Delta X|$ as a measure of $\sigma$, the straight line with slope equals to 2 in Fig.~\ref{1} can be also understood as the case of the log-Brownian motion but for the returns increments itself. This could be seen as a benchmark and again stresses the need of including memory effects in any market model for getting the appropriate curves from levels greater and lower than the normal level.

We also note that the range where statistics is reliable enough appears to be between 0.01 and 10 times the normal level. The index with less statistics is the CAC-40 which is in fact the market with a lower data amount. The most reliable data statistics is that of the Dow Jones which allows us to look at data below 0.01 and far beyond 10 times its normal level.

\begin{figure}
\begin{center}
\includegraphics[width=0.45\textwidth,keepaspectratio=true]{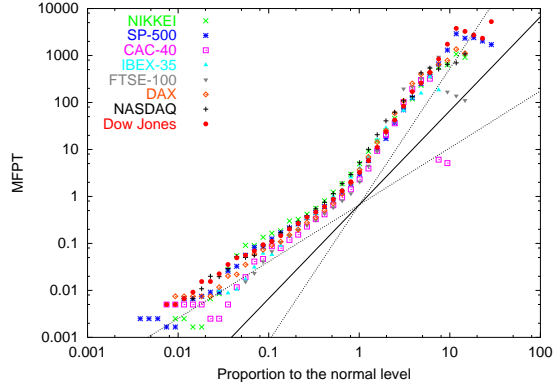} 
\caption{Log-log representation of the empirical MFPT, $\overline{T}(L)$, for the indices outlined in Table I, in terms of the dimensionless critical level $L$. Observe that the MFPT has a very similar behavior for all markets. The solid line represents the Wiener process for the volatility, that is, taking $f(\sigma)=0$ and $g(\sigma)=k$, and for which $\overline{T}(L)=(2/3)L^2$ (as given by Eq.~(\ref{4-5b})). Schematically we can also fit two very distinct regimes for $L<1$ and $L>1$ with $\overline{T}(L)=(2/3)L^\alpha$ whose exponent is respectively $1.21\pm 0.04$ and $2.9\pm 0.1$.}
\label{fig1}
\end{center}
\end{figure}

\begin{figure}
\begin{center}
\includegraphics[width=0.45\textwidth,keepaspectratio=true]{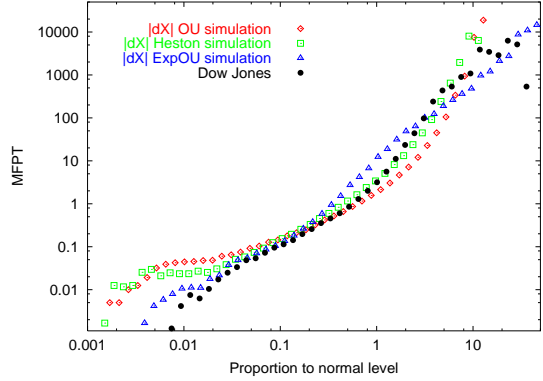} 
\caption{Log-log representation of the empirical MFPT, $\overline{T}(L)$, of the DJIA, along with the OU, Heston and expOU MFPT's. Parameters of the models are the ones provided in the literature~\cite{perello2,yakov,perello3}.}
\label{fig2a}
\end{center}
\end{figure}

\begin{figure}
\begin{center}
\includegraphics[width=0.45\textwidth,keepaspectratio=true]{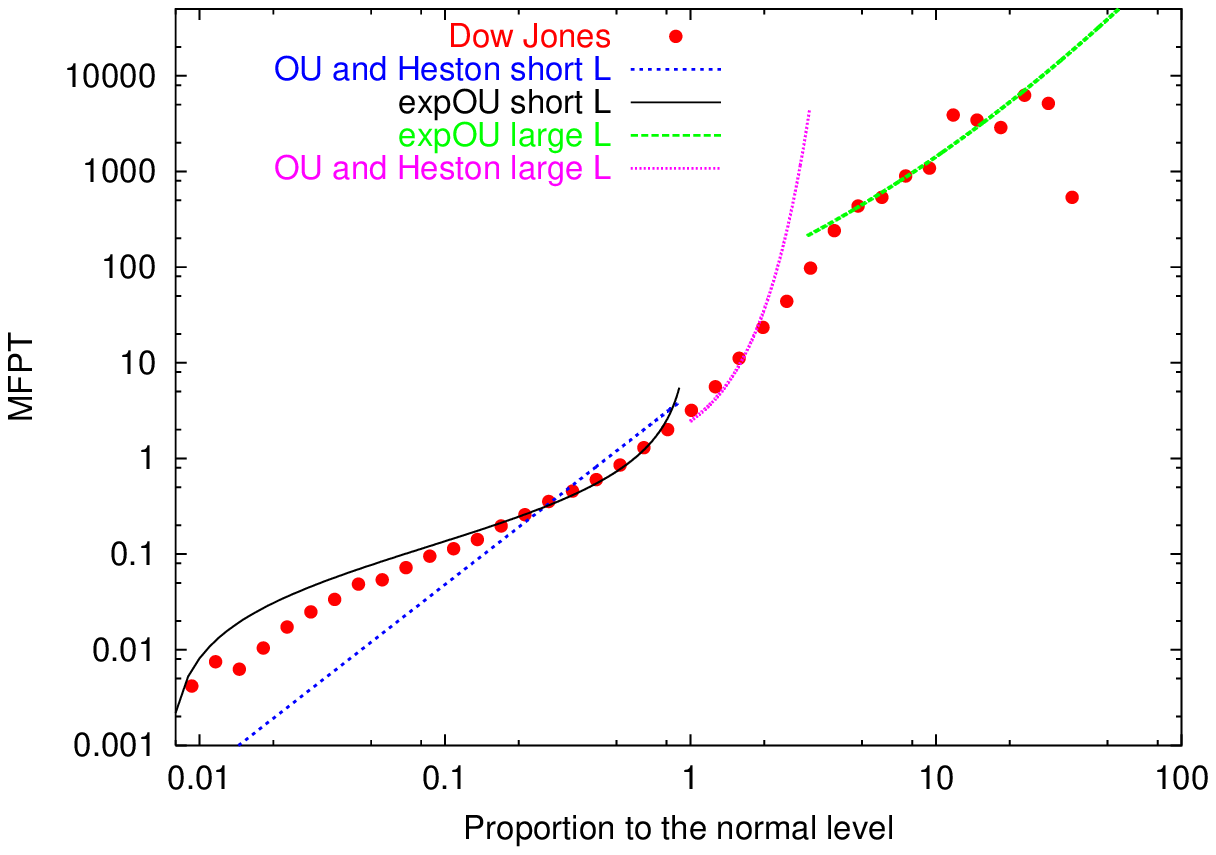} 
\caption{Log-log representation of the empirical MFPT, $\overline{T}(L)$, of the DJIA, along with several asymptotic adjustments corresponding to the theoretical SV models discussed in the text.}
\label{fig2}
\end{center}
\end{figure}

\begin{figure}
\begin{center}
\includegraphics[width=0.45\textwidth,keepaspectratio=true]{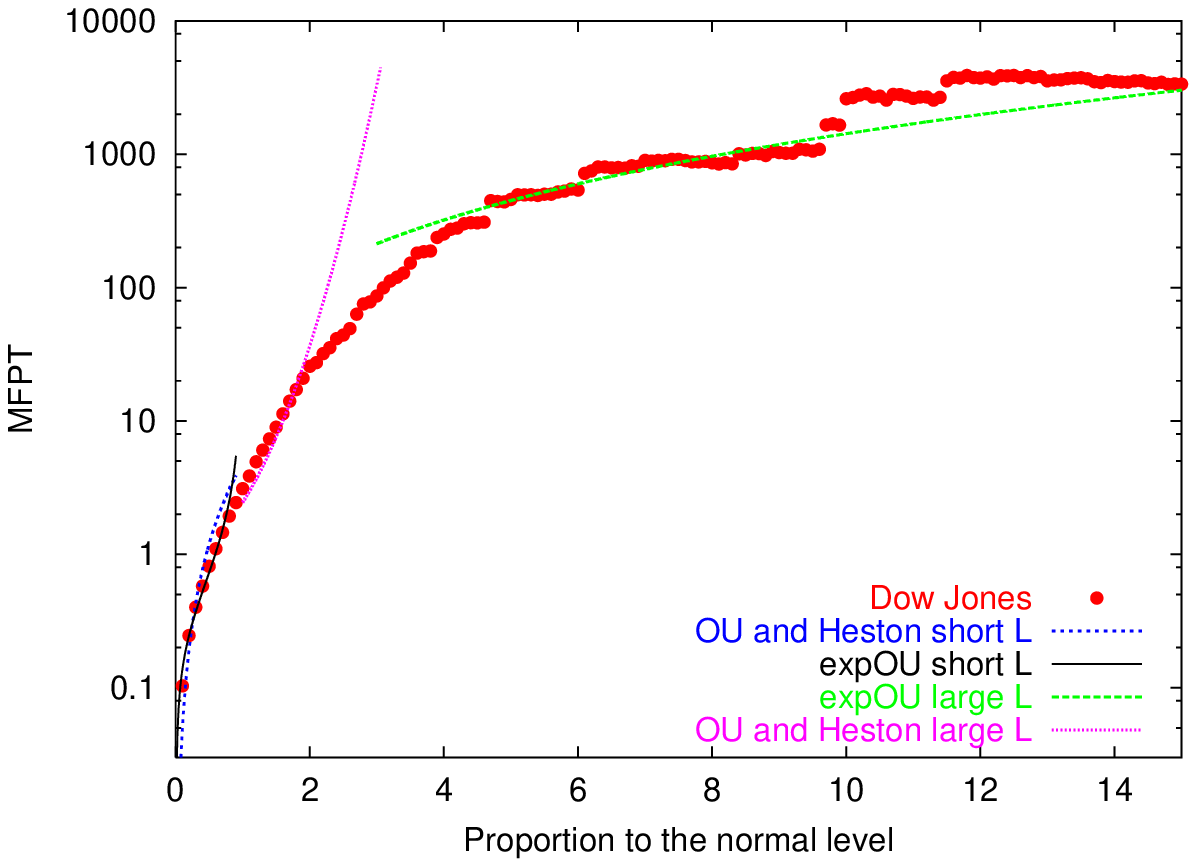} 
\caption{Semi-log representation of the empirical MFPT, $\overline{T}(L)$, of the DJIA, along with several asymptotic adjustments corresponding to the theoretical SV models discussed in the text.}
\label{fig3}
\end{center}
\end{figure}

\begin{figure}
\begin{center}
\includegraphics[width=0.45\textwidth,keepaspectratio=true]{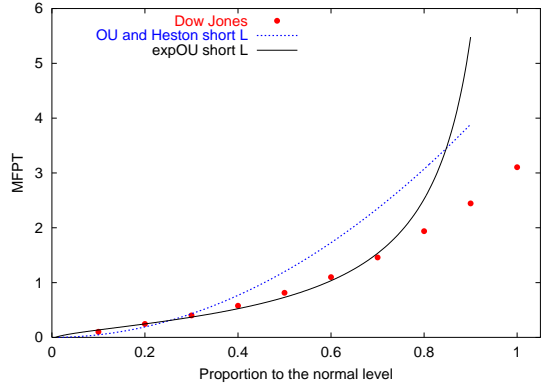} 
\caption{Regular representation of the empirical MFPT of the DJIA for small values of the critical level $L$. The dotted line corresponds to the quadratic behavior $\overline{T}(L)\sim L^2$ shown by OU and CIR-Heston models, Eqs.~(\ref{4-14})-~(\ref{4-15}). Solid line corresponds to the logarithmic increase $\overline{T}(L)\sim 1/\ln L$ of the ExpOU model, Eq.~(\ref{4-16}).}
\label{fig4}
\end{center}
\end{figure}

\begin{figure}
\begin{center}
\includegraphics[width=0.45\textwidth,keepaspectratio=true]{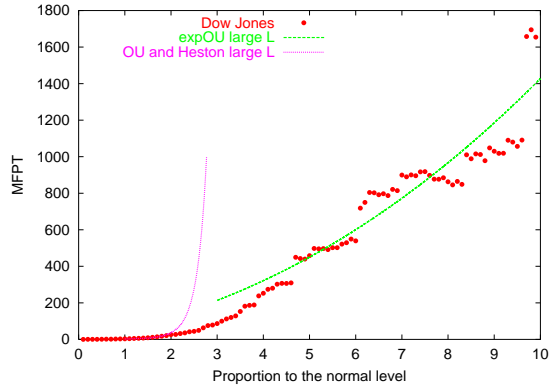} 
\caption{Regular representation of the empirical MFPT of the DJIA for a wide range of values $L$. The dotted line corresponds to the quadratic exponential growth, $\overline{T}(L)\sim e^{L^2}$, shown by OU and CIR-Heston models, Eqs.~(\ref{4-23})-~(\ref{4-24}). Solid line corresponds to the milder exponential growth $\overline{T}(L)\sim e^{\ln^2L}$ of the ExpOU model, Eq.~(\ref{4-25}).}
\label{fig5}
\end{center}
\end{figure}

\begin{table*}
\begin{tabular}{|c|c|c|c|} \hline
Asymptotics & Fitting Region & OU / CIR-Heston & ExpOU \\
\hline
$L\ll 1$ &$0<L\leq 0.4$ & $aL^2$ & $-b/\ln L$
\\
& (25 points) & $a=4.8\pm 0.2 \mbox{ day}$ & $b=0.59\pm 0.02 \mbox{ day}$ \\
\hline
$L\gg 1$ & $1\leq L\leq 10$ & $cL\exp(uL^2)$ & $vL\exp(w\ln^2L)$\\
& (91 points)& $c=1 \mbox{ day}$ & $v=57\pm 1 \mbox{ day}$\\
& & $u=0.9$ & $w=0.17\pm 0.02$\\
\hline
\end{tabular}
\caption{Fits for each SV model in the asymptotic regimes as given by the set of Eqs.~(\ref{4-14a})--(\ref{4-25}). We use a nonlinear least-squares algorithm except from the OU and CIR-Heston case for large $L$ where it is not possible to provide a good fit and we just give some reasonable numbers.}
\label{table3}
\end{table*} 

In the next figures, we compare the experimental result corresponding to the volatility of the DJIA with the theoretical models and their asymptotic expressions. Figure~\ref{fig2a} compares empirical data with simulations taking realistic parameters obtained in the literature~\cite{perello2,yakov,perello3}. We observe that the asymptotic regimes are better described for the expOU than the OU and Heston models. We can go a bit further but now taking the analytical expressions for $\sigma$ provided in previous sections. In Fig.~\ref{fig2}, we can see there that for both small and large critical values, the theoretical model that seems to more accurately follow empirical data is the ExpOU model. The same conclusion is supported by the semi-log representation given in Fig.~(\ref{fig3}). For $L\ll 1$ this is enhanced in Fig.~\ref{fig4} where, in a regular plot, we see that the ExpOU model follows more closely and for longer values of $L$ the empirical result. A similar situation is shown in Fig.~\ref{fig5} where the exponential growth of $\overline{T}(L)$ provided by OU and CIR-Heston models (Eqs.~\ref{4-23}) and~(\ref{4-24})) deviates very quickly from the empirical MFPT, while the slower exponential growth of the ExpOU model (Eq.~(\ref{4-25})) seems to better fit the empirical result. In Table \ref{table3} we show some details of the fitting procedure used to generate Figs. \ref{fig2}--\ref{fig5}. 

\section{Summary and conclusions}
\label{sec6}

We have studied an aspect of the volatility process which is closely related to risk management: its extreme times. The techniques employed have been mainly used in the context of physical sciences and engineering. In this way we have tried to broaden the field of applicability of the mean first-passage time (MFPT) to financial times series. 

We have estimated the empirical MFPT of several major financial indices. For all markets --from the American DJIA to the French CAC-40 and also for different periods of time (see Table~\ref{table1})-- the empirical MFPT follows an universal pattern as shown in Fig.~\ref{fig1}. 

These results sustain the assumed general random character of the volatility similarly to the random diffusion framework and they detect a very different behavior depending on whether the critical level is higher or lower than the average stationary volatility, {\it i.e.,} the volatility's normal level. 

We have found that the MFPT versus the critical level $L$ can be represented as a power law for each regime (see Eq. (\ref{power-law}) and Fig. \ref{fig1}). Compared to the Wiener process in which $\alpha=2$, when the critical level is below the volatility's normal level we get $\alpha \sim 1.2$ which indicates a slower growth of the MFPT than that of the Wiener process. On the other hand, when the critical level is above the normal level the observed exponent is $\alpha \sim 2.9$ and the growth is slower. All of this, has led us to look for theoretical models of the volatility which contain long correlations and are able to describe these distinct patterns depending on the average volatility.

A second issue of our work has been devoted to the most common SV models whose framework is analogous to that of the random diffusion approach. We have therefore chosen the OU, the CIR-Heston and the ExpOU models and obtained closed expressions (up to a quadrature) for the MFPT , $T_\lambda(\sigma)$, to a certain critical level $\lambda$, being $\sigma$ the current value of the volatility (Eqs.~(\ref{3-13}),~(\ref{3-17}) and~(\ref{3-21})). By averaging out the current value of the volatility we have also obtained the averaged MFPT, $\overline{T}(L)$, where $L$ is a dimensionless critical level representing the proportion to the normal level of volatility (Eqs.~(\ref{4-7})-(\ref{4-12})). 

Obviously different SV models furnish different expressions for the MFPT. However, the expressions obtained from OU and CIR-Heston models show a similar behavior while that of the ExpOU model is distinctive. This is clearly seen by asymptotic analysis. Thus, both OU and CIR-Heston models present, for small critical levels, a parabolic increase (Eqs.~(\ref{4-14})-(\ref{4-15})):
$$
\overline{T}(L)\sim L^2, \qquad (L\ll 1);
$$
while for large values of $L$ they show an ``explosive'' quadratic exponential growth (Eqs.~(\ref{4-23})-~(\ref{4-24})):
$$ 
\overline{T}(L)\sim e^{L^2}, \qquad (L\gg 1).
$$

On the other hand the ExpOU model displays, for small critical levels, a logarithmic increase (Eq.~(\ref{4-16})):
$$
\overline{T}(L)\sim 1/\ln L, \qquad (L\ll 1);
$$
and a milder exponential growth when $L$ is large (Eq.~(\ref{4-25})):
$$
\overline{T}(L)\sim e^{\ln^2L}, \qquad (L\gg 1).
$$

Moreover, we have also fitted the above asymptotic approximations provided by SV models to the empirical data (see Figs.~\ref{fig2}-\ref{fig5}), with the overall conclusion that the ExpOU model better explains the experimental MFPT than OU and CIR-Heston models, specially for large values of the critical level. 

\acknowledgments 
The authors acknowledge support from Direcci\'on General de Investigaci\'on under contract No. FIS2006-05204.

\appendix

\section{The normal level of volatility}
\label{appendixA}

We know that the normal level of volatility, $\sigma_{\rm s}$, is defined as the mean stationary value of the volatility process $\sigma(t)$:
\begin{equation}
\sigma_{\rm s}=\int_{-\infty}^{\infty}\sigma p_{\rm st}(\sigma)d\sigma,
\label{a1}
\end{equation}
where $p_{\rm st}(\sigma)$ is the stationary pdf of $\sigma(t)$. For SV models the volatility process is determined by the one-dimensional diffusion given in Eq.~(\ref{3-1}), in this case the stationary distribution is explicitly given by~\cite{gardiner}
\begin{equation} 
p_{\rm st}(\sigma)=\frac{N}{g^2(\sigma)}\exp\left[2\int\frac{f(\sigma)}{g^2(\sigma)}d\sigma\right],
\label{a2}
\end{equation}
where $N$ is a normalization constant. For the SV models discussed in the main text, the stationary pdf reads

(a) {\it OU model}:
\begin{equation}
p_{\rm st}(\sigma)=\frac{\nu}{\sqrt{\pi}}e^{-\nu^2(\sigma-m)^2},
\label{a3}
\end{equation}
where $\nu$ is defined in Eq.~(\ref{3-11}).

(b) {\it CIR-Heston model}:
\begin{equation}
p_{\rm st}(\sigma)=\frac{2^{1+\beta}\nu^{2\beta}}{\Gamma(\beta)}\sigma^{2\beta-1}e^{-2\nu^2\sigma^2},
\label{a4}
\end{equation}
where $\beta$ is given by Eq.~(\ref{3-16}).

(c) {\it ExpOU model}:
\begin{equation}
p_{\rm st}(\sigma)=\frac{\nu Me^{-1/4\nu^2}}{\sqrt{\pi}\sigma^2}e^{-\nu^2\ln^2(\sigma/M)},
\label{a5}
\end{equation}
where $M$ is defined in Eq.~(\ref{M}). 

Finally from Eqs.~(\ref{a1}) and Eqs.~(\ref{a3})-~(\ref{a5}) we immediately obtain the normal level of every SV model ({\it cf.} Eq.~(\ref{normal-level}):
\begin{equation}
\sigma_{\rm s}=\begin{cases} m & \text{OU model,} \\
 \gamma/\nu\sqrt{2} & \text{CIR-Heston model,} \\
 me^{1/4\nu^2} & \text{ExpOU model.}
 \end{cases}
\label{a6} 
\end{equation} 

\section{Asymptotic approximations}
\label{appendixB} 

In this appendix we will prove that a convenient asymptotic representation of the averaged MFPT, for large values of the critical level, is given by Eq.~(\ref{4-22}). 

The starting point of our derivation is the general expression of the MFPT given by Eq.~(\ref{3-8}). We introduce this expression into the definition of the averaged MFPT given in Eq.~(\ref{4-1}) and then exchange the order in which the double integral is performed. We have
\begin{equation}
\overline{T}(\lambda)=\frac{2}{\lambda}\int_0^\lambda \frac{e^{\psi(x)}}{g^2(x)}dx\int_x^\lambda ye^{-\psi(y)}dy,
\label{b1}
\end{equation}
which, after defining
\begin{equation}
h(x,\lambda)\equiv\frac{1}{g^2(x)}\int_x^\lambda ye^{-\psi(y)}dy,
\label{b2}
\end{equation}
can be written as
\begin{equation}
\overline{T}(\lambda)=\frac{2}{\lambda}\int_0^\lambda h(x,\lambda)e^{\psi(x)}dx.
\label{b3}
\end{equation}

Let us now see that, in all SV models studied in this paper, the function 
$$
\chi(x)=e^{\psi(x)}
$$
reaches its maximum value at a point $x_m$ which, for $\lambda$ sufficiently large, is inside the integration interval $(0,\lambda)$ of Eq.~(\ref{b3}). Indeed, the extremes of $\chi(x)$ coincide with extremes of $\psi(x)$, and from Eq.~(\ref{3-9}) we see that the extreme points of $\psi(x)$ are those in which the drift $f(x)=0$ vanishes. Therefore, for all SV models here analyzed, there is one and only one extreme given by 
\begin{equation}
x_m=\begin{cases} m & \text{OU model,}\\
\sqrt{m} & \text{CIR-Heston model,}\\
M & \text{ExpOU model.}
\end{cases}
\label{b4} 
\end{equation} 
Note that for $\lambda$ large enough (in fact larger than the maximum value of $m$, $\sqrt{m}$ and $M$) $x_m$ lies inside the interval $(0,\lambda)$. Moreover, $x_m$ is a maximum since the second derivative $\psi''(x)=2f'(x)/g^2(x)<0$ is negative for all models. 

\begin{figure}[htb]
\begin{center}
\includegraphics[width=0.45\textwidth,keepaspectratio=true]{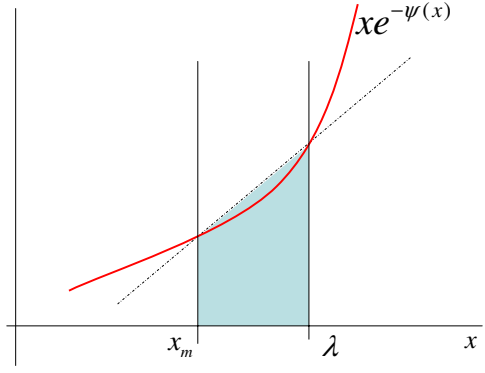} 
\caption{Approximate evaluation of $h(x_m,\lambda)$.}
\label{fig6}
\end{center}
\end{figure}

The fact that the function $\chi(x)=e^{\psi(x)}$ reaches its maximum value inside the integration interval $(0,\lambda)$ of Eq.~(\ref{b3}) allows us to apply Laplace's method for the approximate evaluation of the integral~\cite{erderlyi}. Expanding $\psi(x)$ around $x_m$
$$
\psi(x)=\psi(x_m)+\psi''(x_m)x^2/2+O(x^3),
$$
and substituting this into Eq.~(\ref{b3}) we get
$$
\overline{T}(\lambda)\sim\frac{2}{\lambda}h(x_m,\lambda)e^{\psi(x_m)}\int_0^\lambda e^{\psi''(x_m)x^2/2}dx,
$$
but it is supposed that $e^{x^2\psi''(x_m)/2}$ falls off quickly to zero. Hence, for sufficiently large values of $\lambda$, we may write
\begin{widetext}
$$
\int_0^\lambda e^{\psi''(x_m)x^2/2}dx\simeq\int_0^\infty e^{\psi''(x_m)x^2/2}dx
=\frac{\sqrt{\pi}}{2\sqrt{-\psi''(x_m)}},
$$
\end{widetext}
whence
\begin{equation}
\overline{T}(\lambda)\sim\frac{\sqrt{2\pi}}{\lambda\sqrt{-\psi''(x_m)}}h(x_m,\lambda)e^{\psi(x_m)}.
\label{b5}
\end{equation}

Now, in order to obtain an asymptotic approximation of $\overline{T}(\lambda)$ as $\lambda\rightarrow\infty$, we have to discern the behavior of the quantity ({\it cf} Eq. (\ref{b2})):
$$
h(x_m,\lambda)=\frac{1}{g^2(x_m)}\int_{x_m}^\lambda xe^{-\psi(x)}dx
$$
as $\lambda$ becomes large. One can easily show that the integrand of this equation, $xe^{-\psi(x)}$, is an increasing function of $x$ for 
$x\geq x_m$ (see Fig.~\ref{fig6}). We can, therefore, approximate the integral by the value of the area of the trapezoid depicted in Fig.~\ref{fig6} with the result
$$
h(x_m,\lambda)\simeq \frac{1}{2g^2(x_m)}(\lambda-x_m)\left[\lambda e^{-\psi(\lambda)}+x_me^{-\psi(x_m)}\right],
$$
and for large values of $\lambda$ we can write
\begin{equation}
h(x_m,\lambda)\simeq \frac{1}{2g^2(x_m)}\lambda^2 e^{-\psi(\lambda)},
\label{b6}
\end{equation}
which is the approximation sought for $h(x_m,\lambda)$. Substituting this into Eq.~(\ref{b5}) and taking into account that $|\psi(x_m)|\ll|\psi(\lambda)|$ ($\lambda\rightarrow\infty$) we finally prove Eq.~(\ref{4-22}):
\begin{equation}
\overline{T}(\lambda)\sim\frac{\lambda\sqrt{\pi}}{g^2(x_m)\sqrt{-2\psi''(x_m)}}e^{-\psi(\lambda)}.
\label{b7}
\end{equation}


\begin{thebibliography}{99}
\bibitem{footnote1} Had we taken the logarithmic return the mathematical analysis would be model dependent and more complicated. Indeed, by 
considering It\^o lemma we have
$$
d(\ln S(t))={dS}/{S}-(\sigma^2/2)dt,
$$
which depends on the volatility and hence on the SV model chosen. We have nonetheless compared the results for the MFPT using these two ways of defining the return and we have not seen any significant difference.
\bibitem{footnote2} This procedure of extracting $\sigma$ from data furnishes only a ``proxy'' of the true value of the volatility. There might be more accurate analyses to obtain this value but any of these methods needs to assume an specific SV model which is one thing we want to avoid. In any case, the effort is far beyond the scope of this paper and we leave it for future future research.
\end{thebibliography}
\end{document}